\documentclass[11pt,a4paper]{article}
\usepackage[utf8]{inputenc}
\usepackage[T1]{fontenc}
\usepackage{lmodern}
\usepackage{microtype}
\usepackage[margin=2.5cm]{geometry}
\usepackage{setspace}
\usepackage{parskip}
\usepackage{graphicx}
\usepackage{booktabs}
\usepackage{enumitem}
\usepackage{hyperref}
\hypersetup{colorlinks=true,linkcolor=blue,citecolor=blue,urlcolor=blue}
\usepackage{titling}
\usepackage{fancyhdr,natbib}
\usepackage{caption}
\usepackage{amsmath,amssymb,comment}

\pretitle{\vspace*{-1cm}\begin{center}\LARGE\bfseries}
\posttitle{\par\end{center}\vskip 0.5em}
\preauthor{\begin{center}\large}
\postauthor{\end{center}}
\predate{\begin{center}\large}
\postdate{\end{center}}

\newcommand{\fighere}{%
\begin{figure}[h]
    \centering
    \begin{minipage}[t]{0.46\linewidth} 
    \includegraphics[width=\linewidth]{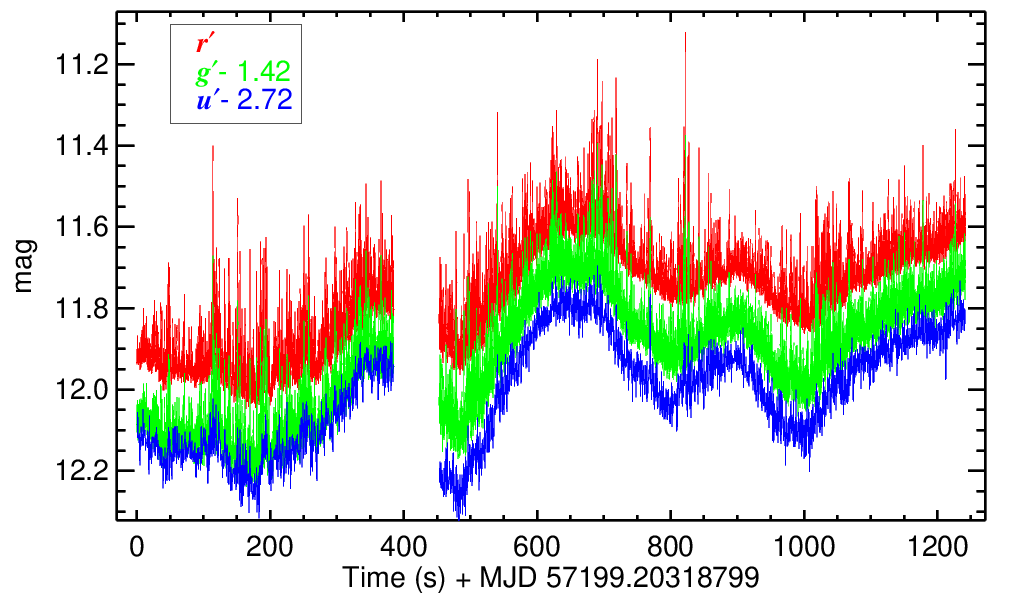}
    \end{minipage}\hfill
    \begin{minipage}[b]{0.47\linewidth}
    \vspace{-0.2em}
    \captionof{figure}{\footnotesize
    {\bf Example of stochastic variability in $ugr$ filters for the accreting black hole X-ray binary V404\,Cygni. Flux variations span a wide range of Fourier frequencies from milliseconds to hours. Statistical uncertainties are too small to discern. Variability strength changes from $u$ to $r$, allowing the underlying non-thermal processes in the vicinity of the compact object to be disentangled.~\\}}
    \label{fig:example}
    \end{minipage}
\end{figure}
}

\begin{document}

\begin{titlepage}
  \centering
  \vspace*{1cm}
  {\Huge\bfseries Expanding Horizons \\[6pt] \Large Transforming Astronomy in the 2040s \par}
  \vspace{1.5cm}

  {\LARGE \textbf{Accretion and Ejection Physics at High Time Resolution}\par}
  \vspace{1cm}

  \begin{tabular}{p{4.5cm}p{10cm}}
    \textbf{Scientific Categories:} & Time-domain, Stars, Binaries, Accretion\\
    \\
    \textbf{Submitting Author:} & Federico Vincentelli: \\
    & Affiliation:  Fluid and Complex Systems Centre, Coventry University (UK) \\
    & Email: federico.vincentelli@coventry.ac.uk\\
    \\
    \textbf{Contributing authors:} 
& Piergiorgio Casella (INAF-OAR, IT) \\
& Alexandra Veledina (University of Turku, FI) \\
& Alessandra Ambrifi (IAC, ES) \\
& Maria Cristina Baglio (INAF-OAB, IT) \\
& David Buckley (SAAO, ZA) \\
& Noel Castro Segura (University of Warwick, UK)\\ 
& Yuri Cavecchi (UPC, ES) \\
& Domitilla de Martino (INAF-OACN, IT) \\
& Melania Del Santo (INAF-IASF Palermo, IT) \\
& Poshak Gandhi (University of Southampton, UK) \\
& Giulia Illiano (INAF-OAB, IT) \\
& Riccardo La Placa (INAF-OAR, IT)\\
& Christian Malacaria (INAF-OAR, IT) \\
& Alessio Marino (ICE-CSIC, ES) \\
& Kieran O'Biren (Durham University, UK) \\
& Nanda Rea (ICE-CSIC, ES) \\
& Andrea Sanna (University of Cagliari, IT) \\
& Simone Scaringi (Durham University, UK) \\
& Tariq Shahbaz (IAC, ES) \\
& Luca Zampieri (INAF-OAPd, IT) \\
\end{tabular}

  \vspace{1cm}

\end{titlepage}


\section{Introduction and Background}
\label{sec:intro}

 Accretion discs are ubiquitous in the Universe and are the central engine of a large fraction of high-energy astrophysical sources spanning a huge range of masses, from Cataclysmic Variables to Active Galatic Nuclei [1]. Notably, discs not only produce X-ray radiation but also launch powerful outflows, such as equatorial winds [2] or collimated jets [3], which have a significant impact on both their hosting systems and the surrounding environment [4, 5].
Despite decades of studies across different wavelengths, many questions about the structure and evolution of these systems remain unanswered. Furthermore, it is still unclear how the strong outflows are produced during the accretion events. Such a lack of knowledge limits our understanding of some of the most extreme phenomena in the Universe, and prevents us from thoroughly assessing the role of accretion on larger scales [6].

Among the different accreting systems, low-mass X-ray binaries (LMXBs) display the ideal properties to answer the main open problems regarding accretion: they contain either a stellar mass black hole (BH) or a neutron star (NS) accreting from a low-mass stellar companion. Most notably, they are visible over a broad energy range [7], with every band providing information about a specific step in the accretion/ejection flow. While the high-energy spectrum (UV to hard X-rays) is dominated by the emission from the inner disc, the lowest energies arise from the jets (radio to infrared). Crucially, the optical-infrared (O-IR) band emerges from the intersection of these two components and is key to understanding how inflows are converted into outflows and the specific geometry of the innermost regions where this conversion occurs. The O-IR range also provides strong emission and absorption lines,  coupled with blue-shifted absorptions, a key signature of equatorial winds [8].


A fundamental property of accretion is the ubiquitous presence of stochastic variability across the electromagnetic spectrum [9, 10] (see Fig.~\ref{fig:example}). Understanding how this variability originates and propagates among the different emitting components is a powerful way to study their structure and connections. Crucially, LMXBs exhibit variability ranging from tens of seconds down to sub-second timescales. The recent expansion of high-time-resolution O-IR instruments - with key contributions from ESO - has opened a new window onto this regime, revealing a wealth of previously inaccessible phenomenology [11, 12] related to the accretion-ejection processes. In the following paragraphs we will highlight the most significant ones:

\fighere{}

%


\textbf{Physical processes within magnetised plasmas: jets and accretion flow.} Since the early detection of O-IR variability on sub-second timescales, it has been clear that such rapid variations must arise from synchrotron emission. The discovery of a short (100~ms) lag between the X-ray and O-IR variable emission in BH LMXBs during a jet-dominated state [11, 12] clearly showed that such lag was the propagation time from the inner accretion flow to the base of the jet.  \looseness=-3

Such a measurement opened a new window on the internal processes of these outflows. By extending a well-known idea in the Gamma-ray burst field[13], it was possible to show that the spectral and variability properties observed in LMXBs could be explained with accretion-driven shocks within the jet . Detailed modelling of the O-IR timing properties allowed, for the first time, constraints to be placed on the geometry of the jet and its speed.

Intriguingly, the connection between the X-ray and the O-IR seems to change from source to source and throughout the outburst. Observations have shown that X-ray and O-IR variability can also be anti-correlated, pointing towards a scenario in which the emitting components share the same energy budget [14]. Thanks to a larger data sample, it was possible to show that this feature is more likely to arise from self-synchrotron Comptonization in an extended magnetised corona [15]. The presence of this component has substantial consequences for the structure of the accretion flow, implying that it is most likely stratified and threaded by a magnetic field. Yet, due to the lack of intensive monitoring, it is still unclear when or why jet or magnetised hot flow can be more dominant in the O-IR range. 

\textbf{Structure of the innermost regions of accreting systems.} One of the unique features of LMXBs is the appearance of X-ray quasi-periodic oscillations (QPOs), which vary in frequency during their outbursts from $\approx$0.1 Hz upto $\approx$10 Hz[16]. 
Theoretical calculations indicate that such frequencies can be explained by precession of the inner disc, offering strong constraints on the geometry of these systems[17]. Notably, O-IR QPOs have also been detected, sometimes at half or equal to the X-ray QPO frequencies, or even without a significant X-ray counterpart [11,18]. However, due to the small amount of data, their multi-wavelength behaviour remains unclear. Both the jet and the hot flow can, in principle, produce O-IR QPOs, albeit with model-dependent limitations [18,19]. As the frequency of these oscillations strongly depends on the height and radius of the precessing medium [18], understanding this behavior would provide crucial information on the structure of the flow around the compact object.

\vspace{-0.2cm}
\section{Open Science Questions in the 2040s}
\label{sec:openquestions}


In the next decade, new facilities as the LSST (optical), SKA (radio) and eXTP (X-rays), along with with improved 3D relativistic simulations, will greatly refine our understanding of accretion discs, jets, and equatorial winds. Yet, without high-quality timing data, several fundamental questions will remain open. Two of the most significant one are outlined below.\looseness=-10


\textbf{What is the structure of the innermost regions of these systems? What is the radial profile of the outflows?} 

Fast optical-infrared variability has enabled us to probe the processes at the base of the jet and study its dynamics. Yet this is still very limited.  While some constraints have been put, they are still highly dependent on simplified assumptions and models. As general relativistic 3D simulations improve over the coming years, it will be essential to obtain more stringent constraints, especially on the physics/processes at the interface between the disc and the outflow. These are, for example, what is the radial structure of the jet? Where are equatorial outflows launched within the disc? What triggers the formation and the structure of these outflows?

Answering these questions is key not only to fully understand the observed phenomena, but especially to quantify the impact of accretion onto larger scales. To tackle them, it is necessary to decompose all the physical components in time and wavelength adequately, and to link them to the foreseen high-resolution X-ray data available in the following decades. The only way to do this is to have a facility that provides both high-resolution spectroscopy and time resolution.

\textbf{Is the accretion-ejection process the same in black holes and neutron stars?}

Decades of X-ray spectral timing and radio studies revealed unambiguously a common picture for accretion states between BHs and NSs. Fast O-IR variability can therefore be used to quantify how universal accretion is. So far, the vast majority of data comes from BH systems, thus biasing, in part, our knowledge of accretion-ejection processes toward these systems. By studying the similarities and differences between BH and NS systems, future multi-wavelength timing analysis can be used to tackle fundamental questions, such as quantifying the effects of a hard surface (for NSs) versus an event horizon (for BHs) or a strong magnetic field.\looseness=-9

Today we only have a few examples of stochastic fast non-thermal variability in NSs. However, are still limited by the current time resolution of IR imaging instruments in the Southern hemisphere($\approx$0.125\,s) [20,21]. As NSs show the bulk of their variability up to $\approx$100-1000 Hz, to perform a meaningful analysis of these systems we need to probe much higher temporal frequencies. Today, there is no publicly available imaging or spectroscopic instrument that can reach such time resolution. Furthermore, it is unclear whether such an instrument will be developed in the near future.

\vspace{-0.2cm}
\section{Technology and Data Handling Requirements}
\label{sec:tech}

To answer these key questions on accretion and to isolate the relevant physical components with sufficient spectral and temporal resolution, several technological developments are required. These will enable us to fully resolve the behavior of the inflow--outflow structure.


\textbf{Wavelength dependence:}
Strictly simultaneous, high-cadence optical–infrared observations are essential for mapping the accretion flow. Covering the UV (400~nm) to IR ($>$2000~nm) range is particularly important, as different flow geometries (e.g., jet stratification or disc temperature gradients) produce measurable inter-band time delays. Recent optical experiments (400–1000~nm) show delays as small as $\approx$33 $\mu$s nm$^{-1}$ [22].  Resolving key spectral features—such as P-Cygni profiles from disc winds—requires a resolution of at least
R$>$5000 would be necessary to separate continuum and absorption while preserving timing accuracy.

\textbf{Photon Counting}
Accessing variability above 100\, Hz — especially in NSs, where most of the power lies at high frequencies — requires true photon-counting detectors capable of time-tagging individual photons with microsecond precision. Unlike CCD/CMOS systems, which are fundamentally limited by read-out noise and minimum integration time, photon-counting architectures operate with effectively zero read-out noise. This enables high-cadence measurements even at very low fluxes. Such detectors are therefore essential for a next-generation timing facility aimed at revealing the fastest processes in compact objects.

\textbf{Flexible Scheduling}
Despite the power of fast multi-wavelength timing, progress is currently constrained by the absence of a dedicated facility. Strong stochastic variability in Galactic transients usually appears only during brief 1–2-week intervals at the start and end of outbursts, requiring dense, rapid-response monitoring. Oversubscription and rigid scheduling often prevent coverage of these critical windows. A network of small- to medium-class telescopes with dynamic scheduling would provide the flexibility needed to capture these short-lived but scientifically crucial events.\looseness=-3

{\footnotesize\textbf{References:} 
[1]~Pringle 1981, ARA\&A, 19, 137
[2]~Muñoz-Darias + 2016, Nature, 534, 75
[3]~Fender + 2004, MNRAS, 355, 1105
[4]~Corbel \& Fender 2002, ApJL, 573, L35
[5]~Fabian 2012, ARA\&A, 50, 455
[6]~Weinberger + 2017 MNRAS 465, 3291
[7]~Gandhi + 2025 arXiv:2510.01338
[8]~S\'anchez-Sierras \& Mu\~noz-Darias A\&A 640, L3 
[9]~Tetarenko + 2021 504, 3862
[10]~Scaringi + Science Ad. 1,  9, e1500686
[11]~Gandhi + 2010, MNRAS, 407,2166
[12]~Casella + 2010, MNRAS, 404, L21
[13]~Malzac + 2014 MNRAS  443, 299
[14]~Malzac + 2004, MNRAS, 351, 253
[15]~Veledina + 2017, MNRAS, 470, 48
[16]~Motta + 2011, MNRAS, 418
[17]~Ingram \& Motta 2019, NewAR, 85, 101524
[18]~Malzac + 2018, MNRAS, 480, 2054
[19]~Veledina + 2013, ApJ, 778, 165
[20]~Vincentelli + 2021, Nature, 615, 45
[21]~Vincentelli + 2023, MNRAS, 525, 2509
[22]~Vincentelli + 2025 MNRAS 539,2347}

\end{document}